\pdfoutput=1
\documentclass[aps,prl,reprint,superscriptaddress,usenames,dvipsnames]{revtex4-1}

\usepackage[dvipsnames]{xcolor}
\usepackage[utf8]{inputenc}
\usepackage[T1]{fontenc}
\DeclareUnicodeCharacter{2212}{-}

\usepackage{graphicx}
\usepackage{amsmath}
\usepackage{siunitx}
    \DeclareSIUnit\angstrom{\text {Å}}
    \DeclareSIUnit\rydberg{Ry}
\usepackage[hidelinks]{hyperref}
\usepackage{chemformula}
\usepackage[version=4]{mhchem}
\usepackage{soul}

\begin{document}

%\preprint{}

\title{Symmetry Breaking in the Superionic Phase of Silver-Iodide}

\author{Amir Hajibabaei}
\email{ah2281@cam.ac.uk}
\affiliation{Yusuf Hamied Department of Chemistry, University of Cambridge, Cambridge, CB2 1EW, United Kingdom}
\author{William J. Baldwin}
\affiliation{Engineering Laboratory, University of Cambridge, Cambridge, CB2 1PZ, United Kingdom}
\author{G\'abor Cs\'anyi}
\affiliation{Engineering Laboratory, University of Cambridge, Cambridge, CB2 1PZ, United Kingdom}
\author{Stephen J. Cox}
\email{sjc236@cam.ac.uk}
\affiliation{Yusuf Hamied Department of Chemistry, University of Cambridge, Cambridge, CB2 1EW, United Kingdom}

%\thanks{}

\date{\today}

\begin{abstract}
In the superionic phase of silver iodide, we observe a distorted tetragonal structure characterized by symmetry breaking in the cation distribution. 
This phase competes with the well known bcc phase, with a symmetric cation distribution at an energetic cost of only a few meV/atom.
The small energy difference suggests that these competing structures may both be thermally accessible near the superionic transition temperature.
We also find that the distribution of silver ions depends on the low-temperature parent polymorph, with memory persisting in the superionic phase
on the nanosecond time scales accessible in our simulations.
Furthermore, simulations on the order \SI{100}{\nano\second} reveal that even at temperatures where the bcc phase is stable, significant fluctuations toward the tetragonal lattice structure remain.
Our results are consistent with many ``anomalous'' experimental observations and offer a molecular mechanism for the ``memory effect'' in silver iodide.
\end{abstract}
%\keywords{}

\maketitle

%----------------------------------------------------------------------------------------------------------------
% ------------------- 1st paragraph -----------------------------------------------------------------------------
%----------------------------------------------------------------------------------------------------------------
Solid electrolytes offer great promise as materials for energy storage owing to their excellent ionic conductivity, with relatively high energy densities while remaining safe to use.
But the atomic mechanisms that govern their behavior are far from simple~\cite{Famprikis2019FundamentalsBatteries}. 
For example, nanoscale diffusion in superionic conductors can substantially differ from 
the typical Brownian motion~\cite{Song2019TransportFluids,Marcolongo2017IonicElectrolytes,Poletayev2024TheOptics,He2017OriginConductors,NMVargas-Barbosa2020DynamicTransport,Muy2021PhononionDynamics,Gao2020ClassicalConductors}. 
The archetypal type I solid electrolyte used as a model to understand these systems is silver iodide~\cite{Hull2004Superionics:Processes}, yet many of its properties remain poorly understood~\cite{Nield1995Fast-IonIodide}.

%----------------------------------------------------------------------------------------------------------------
% ------------------- 2st paragraph -----------------------------------------------------------------------------
%----------------------------------------------------------------------------------------------------------------
At ambient conditions, AgI assumes hexagonal/cubic close-packed structures with many possible polymorphs
resulting from variations in the stacking sequence.
The main polymorphs with ordered stacking sequences are wurtzite ($\beta$) and
zincblende ($\gamma$)~\cite{Burley1963PolymorphismIodide,Burley1963StructureIodide,Takahashi1969TheIodide}.
Heated above \SI{147}{\celsius}, a $\beta/\gamma$ mixture transitions into the superionic $\alpha$ phase with a bcc \ce{I-} framework.
It is known, from nearly a century ago~\cite{Bloch1931UberJodsilbers}, 
that the $\alpha$ phase can retain a memory of its parent structure
evident from the $\beta/\gamma$ composition obtained upon cooling.
This ``memory effect'' was systematically investigated more recently~\cite{Smith2019StackingPhase}
by heating and cooling samples with well-controlled degrees of stacking disorder,
which showed that the degree of persisting memory 
also depends on kinetic factors such as the cooling rate.
As the $\beta/\gamma \to \alpha$ transition
occurs rapidly and with no remaining traces of the low temperature phases,
mechanisms such as nucleation and crystallization are generally ruled out.

%----------------------------------------------------------------------------------------------------------------
% ------------------- 3nd paragraph -----------------------------------------------------------------------------
%----------------------------------------------------------------------------------------------------------------
Although a clear explanation for this memory effect has remained elusive, there is ample evidence to suggest that the picture painted by the notion of a straightforward $\beta/\gamma\to\alpha$ phase transition is too simplistic~\cite{Bloch1931UberJodsilbers,Smith2019StackingPhase,Burley1967TheIodide}.
For example, it has been proposed that \ce{Ag+} ions preferentially occupy certain sites within the \ce{I-} bcc framework, and that the degree of preference for certain sites depends upon the $\beta/\gamma$ stacking composition of the low-temperature parent phase~\cite{Burley1967TheIodide}.
In this seminal study of the memory effect, Burley proposed that such preferential site occupation was responsible for sample-dependent variations in integrated intensities of the diffraction pattern in the $\alpha$ phase.
Moreover, he also noted that any memory is irreversibly lost when temperatures exceed
\SI{170}{}-\SI{175}{\celsius}.

%###############################################################################################
%############### Figure 1  #####################################################################
%###############################################################################################
\begin{figure*}[t!]
    \centering
    \includegraphics{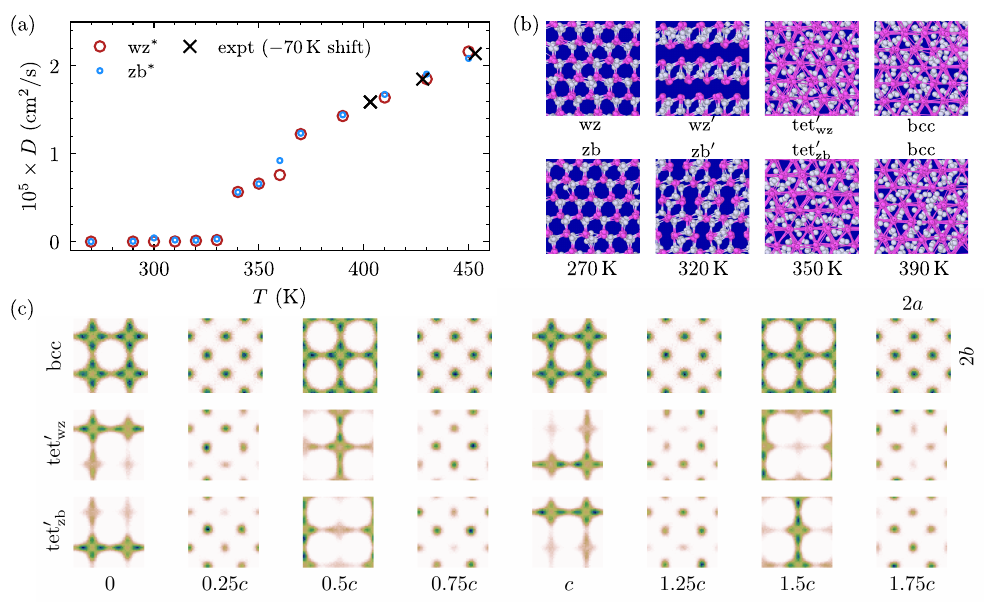}
    \caption{
        \label{fig:md-parents}
        The phase changes of \ce{AgI} at $P=0$\,GPa derived from wurtzite (wz$^\ast$) and zincblende (zb$^\ast$) configurations.
        (a) Self diffusion constants of \ce{Ag+} indicate a superionic transition at $T\approx 340$\,K in the simulations.
        The diffusion constants from simulation are in good agreement with experimental data~\cite{Kvist1970Self-diffusionIodide}, once a \SI{70}{\kelvin} shift is accounted for.
        (b) Snapshots at different temperatures (as indicated) show intermediate structures between wz$^\ast$/zb$^\ast$ and the expected bcc lattice for the superionic phase. Note that only part of the simulation cell is shown for clarity.
        (c) The \ce{Ag+} distributions are shown by cross sections along the $c$-axis.
        For the distorted tetragonal lattices tet$^\prime_{\rm wz}$ and tet$^\prime_{\rm zb}$ at \SI{350}{\kelvin}, the \ce{Ag+} distribution has a broken symmetry which depends on the low temperature parent phase. 
        The site occupation symmetry is recovered in the bcc phase at \SI{390}{\kelvin}.
    }
\end{figure*}

%----------------------------------------------------------------------------------------------------------------
% ------------------- 4rd paragraph -----------------------------------------------------------------------------
%----------------------------------------------------------------------------------------------------------------
Experiments at higher temperatures further demonstrate AgI's complex phase behavior.
For example, at approximately \SI{427}{\celsius}, AgI undergoes a further order-disorder transition, which, in purely stoichiometric samples,
exhibits an anomalous heat capacity~\cite{Perrott1968HeatSamples,Vargas1990AnomalousIodide}.
In a subsequent theoretical analysis, Perrott and Fletcher attribute this observation to entropic changes,
of which the configurational entropy of \ce{Ag+} plays a major role~\cite{Perrott1968HeatTheory}.
Early Raman spectroscopy experiments generally support this scenario~\cite{Fontana1980Raman-AgI,Mariotto1980TemperatureCrystals,Mariotto1981Temperature/math,Mazzacurati1982Theoretical-AgI}.
More recent Raman polarization-orientation measurements~\cite{Brenner2020AnharmonicAgI}
on single crystals of the $\alpha$ phase found crystal-like features that could not be 
accounted for solely by the bcc \ce{I-} host lattice, nor by assuming a crystal-like average distribution of the mobile \ce{Ag+}. 
Instead, this observation was attributed to strongly anharmonic \ce{I-} lattice vibrations
that are coupled to \ce{Ag+} diffusion.

%----------------------------------------------------------------------------------------------------------------
% ------------------- 5th paragraph -----------------------------------------------------------------------------
%----------------------------------------------------------------------------------------------------------------
Molecular simulations are in principle well-placed to provide insight at the microscopic level to help understand such experimental observations.
Indeed, classical molecular dynamics (MD) simulations employing the empirical Parrinello-Rahman-Vashista (PRV)
force field~\cite{Vashishta1978Ionic-AgI,Parrinello1983StructuralConductors}
support the notion of \ce{Ag+} preferentially occupying sites in the
$\alpha$ phase~\cite{OSullivan1991Silver-ionStudy,Madden1992OrderingTransition,Tallon1986Molecular-DynamicsAgI,Tallon1988Constant-stressIodide,Sato2023TopologicalMechanism}.
Despite considerable constraints on the time and length scales that can be probed,
insights from \emph{ab initio} MD (AIMD) simulations~\cite{Wood2006Dynamical/math}
elucidate a dynamic bonding behavior that is challenging to capture with 
conventional empirical force fields.
In particular, recent work has shown that iodide's lone pair electrons, represented by maximally localized Wannier centers,
have a rotational motion that couples to diffusion of \ce{Ag+}~\cite{Dhattarwal2024ElectronicElectrolyte}, 
in an analogous manner to the ``paddle-wheel'' mechanism associated with molecular superionic solids.

%----------------------------------------------------------------------------------------------------------------
% ------------------- 6th paragraph -----------------------------------------------------------------------------
%----------------------------------------------------------------------------------------------------------------
To overcome the limits on accessible time and length scales imposed by AIMD approaches, 
here we exploit recent advances in developing machine learning interatomic potentials (MLIPs) to train 
a surrogate model that represents the underlying potential energy surface (PES)~\cite{Bartok2017, Deringer2021GaussianMolecules},
as determined by the PBE functional~\cite{Perdew1996RationaleApproximations},
with dispersion interactions included using Grimme's D3 correction~\cite{Grimme2010AH-Pu}
with Becke-Johnson damping~\cite{Johnson2005AInteractions}.
Quantum ESPRESSO~\cite{Giannozzi2009QUANTUMMaterials,Giannozzi2017AdvancedESPRESSO,Giannozzi2020QExascale},
along with the pseudopotentials Ag/I.pbe-n-kjpaw-psl.1.0.0.UPF
taken from PSLibrary v1.0.0~\cite{DalCorso2014PseudopotentialsPu},
was used for all DFT calculations.
The DFT settings are discussed further in the Supplemental Materials (SM)~\cite{prl_sm}.

%----------------------------------------------------------------------------------------------------------------
% ------------------- 7th paragraph -----------------------------------------------------------------------------
%----------------------------------------------------------------------------------------------------------------
For the MLIP, we employ the MACE architecture~\cite{Batatia2022MACE:Fields, KovacsMACEeval2023}
to represent the PES.
For training, we employed an active learning procedure with a committee of five MACE models.
The majority of the training data comprises configurations derived from
the wurtzite/zincblende and rocksalt crystal structures by active learning,
though in addition we also included structures derived from other crystal structures found on 
the \emph{Materials Project} online database~\cite{Jain2013Commentary:Innovation}.
In the temperature range of interest for this work, $270\leq T/{\rm K}\leq 450$,
the model demonstrated an accuracy with a root mean squared error of less than ($1$\,meV/atom, $13$\,meV/\AA)
for (energy, force) predictions.
Further information on training and testing are discussed in the SM~\cite{prl_sm}.

%----------------------------------------------------------------------------------------------------------------
% ------------------- 8th paragraph -----------------------------------------------------------------------------
%----------------------------------------------------------------------------------------------------------------
With the trained MLIP, we perform simulations that probe the memory effect. 
Specifically, we use isothermal--isobaric ($NPT$) MD simulations at a pressure $P=0$\,GPa, 
and at a variety of temperatures ($T$), starting from either the wurtzite or zincblende crystal structures, and analyze how the resulting structures depend upon the initial configuration.
The simulation cell comprised twelve $8\times8$ Ag-I double layers with hcp/fcc stacking sequences to construct wurtzite/zincblende structures with 1536 atoms.
Temperature and pressure were maintained with a Nos\'{e}-Hoover thermostat and Parrinello-Rahman barostat, respectively,
as implemented in the Atomic Simulation Environment Python package~\cite{HjorthLarsen2017TheAtoms,Melchionna2000ConstrainedDistribution}.

%----------------------------------------------------------------------------------------------------------------
% ------------------- 9th paragraph -----------------------------------------------------------------------------
%----------------------------------------------------------------------------------------------------------------
Results for the computed self diffusion coefficients $D$ for \ce{Ag+} are presented in Fig.~\ref{fig:md-parents}a. 
For both the wurtzite- and zincblende-derived trajectories, we observe a jump in $D$ at $T\approx 340$\,K, and a second increase at $T\approx 370$\,K.
We also observe that our computed values for $D$ are broadly in line with experiment~\cite{Kvist1970Self-diffusionIodide}, albeit offset by approximately 70\,K~\footnote{These experimental values were obtained not by measuring electrical conductivity, but by tracking changes in the concentration of radioactive silver ions.}. 
Shifting the experimentally observed transition temperature (i.e., $420\,{\rm K}-70\,{\rm K} = 350{\,\rm K}$) by the same amount agrees well with the observed initial jump in $D$.

%----------------------------------------------------------------------------------------------------------------
% ------------------- 10th paragraph -----------------------------------------------------------------------------
%----------------------------------------------------------------------------------------------------------------
Representative snapshots of structures at different temperatures are shown in Fig.~\ref{fig:md-parents}b.
For $T\gtrsim 370$\,K, space group analysis of the average iodide positions confirms the expected bcc framework of the $\alpha$-phase~\cite{Wright1977The-AgI}, with lattice constant $a=5.06$\,\AA{}.
%% \tcb{The variation of $a$ with temperature is also captured consistent with experiment, offset by 55\,K in this case (see SM~\cite{prl_sm})}.
%
For $340\lesssim T/{\rm K} \lesssim 370$, however, we instead observe distorted lattice structures.
The distortions are such that the repeat unit can be described as a $2\times 2\times 2$ superstructure of 
a body-centered tetragonal lattice with space group I4/mmm (see Fig. S3~\cite{prl_sm}).
Although their distortions differ, the  underlying tetragonal lattices of the iodide framework in this temperature range 
are largely similar whether derived from wurtzite or zincblende, with lattice constants $a \approx 4.79$\,\AA{} and $c\approx 5.58$\,\AA{}. 
(Across this temperature range, $a$ and $c$ vary by approx. \SI{0.002}{\angstrom}.)
We refer to such distorted tetragonal structures hereafter as tet$^\prime$, with subscripts to indicate the parent phase when needed (i.e., tet$^\prime_{\rm wz}$ from wurtzite, and tet$^\prime_{\rm zb}$ from zincblende).
At temperatures slightly below, but close to the superionic transition, we observe structures in which the iodide framework is slightly distorted from the underlying parent phase, but with significant rearrangement of the \ce{Ag+} ions.
For wurtzite, these structures are observed for temperatures $320 \lesssim T/{\rm K} \lesssim 330$, while for zincblende they occur across a wider range, $300 \lesssim T/{\rm K} \lesssim 330$.
\ce{Ag+} diffusivity below \SI{340}{\kelvin} is further discussed in the SM~\cite{prl_sm}.

%----------------------------------------------------------------------------------------------------------------
% ------------------- 11th paragraph -----------------------------------------------------------------------------
%----------------------------------------------------------------------------------------------------------------
While the distortions of the iodide frameworks in the tet$^\prime$ structures show relatively minor dependence on the parent phase, 
differences in the distribution of the \ce{Ag+} ions are clear from visual inspection of the trajectories. 
To further analyze this observation, we performed MD simulations of tet$^\prime_{\rm wz}$ and tet$^\prime_{\rm zb}$ at $T = 350$\,K, 
with the volume fixed at the average from the $NPT$ simulations.
We also performed a similar simulation for the bcc structure at $T=390$\,K. The length of these simulations was \SI{4}{}-\SI{6}{\nano\second}.
The calculated histograms of \ce{Ag+} positions are shown in Fig. \ref{fig:md-parents}c
for various slices in the $ab$ plane at different values along the $c$-axis.

%----------------------------------------------------------------------------------------------------------------
% ------------------- 12th paragraph -----------------------------------------------------------------------------
%----------------------------------------------------------------------------------------------------------------

Consistent with the experimental consensus~\cite{Hoshino1977Distribution-AgI,Wright1977The-AgI,Cava1977Single-crystal300C,Keen2002DisorderingConductors},
we observe peaks in the \ce{Ag+} density at the tetrahedral sites [($0,1/2,1/4$) etc.] in the bcc phase, 
with a slight spreading towards the octahedral sites [$(0,1/2,0)$ etc.].
On the other hand, translational symmetry is broken in the tet$'$ phases, 
as reflected in the asymmetric occupation of the tetrahedral sites.
Significant differences between the two become further apparent upon investigation of the connectivity of maxima in the \ce{Ag+} distributions, 
as seen in Fig. S4 in the SM~\cite{prl_sm}.

%###############################################################################################
%############### Figure 2  #####################################################################
%###############################################################################################
\begin{figure}[t!]
    \centering
    \includegraphics{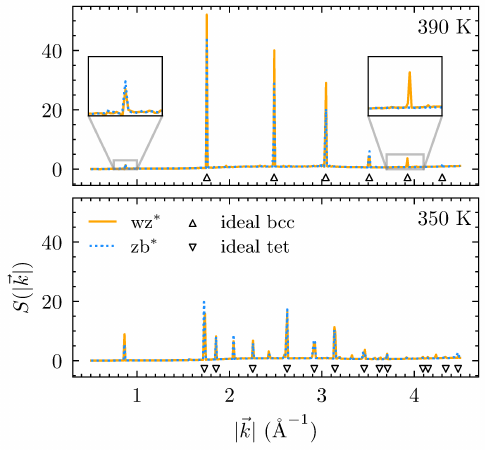}
    \caption{
        \label{fig:sf}
        Memory persists in the superionic phase, as indicated by the structure factors obtained from an average of eight simulations.
        A ``super-structuring'' peak at $\vec{k}\approx 0.87$\,\AA$^{-1}$ is observed in the distorted tetragonal lattices at \SI{350}{\kelvin} (lower panel) which is also observed in the predominantly bcc lattice at \SI{390}{\kelvin} (upper panel). A peak at $\vec{k}\approx 3.93$\,\AA$^{-1}$ is also observed at \SI{390}{\kelvin} for structures derived from wurtize (wz$^\ast)$, which is absent for structures derived from zincblende (zb$^\ast$). In addition, the peak intensities at \SI{390}{\kelvin} differ between the wurtzite and zincblende derived structures. The expected patterns from ideal bcc and tetragonal (with space group I4/mmm) lattices are shown as guide, as indicated by the legend.
    }
    
\end{figure}

%###############################################################################################
%############### Figure 3  #####################################################################
%###############################################################################################
\begin{figure*}[t!]
    \centering
    \includegraphics[width=0.95\textwidth]{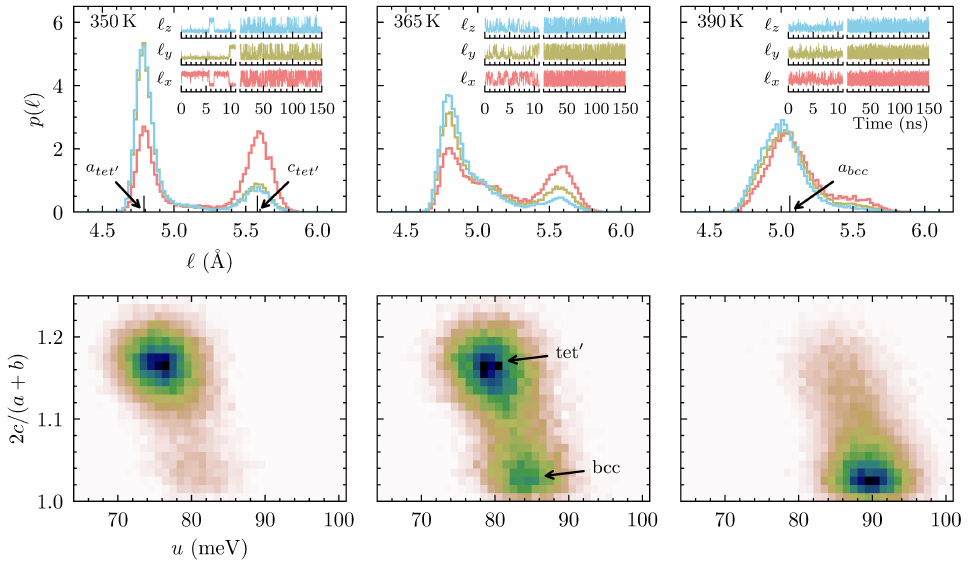}
    \caption{
        \label{fig:rand}
        The distorted tetragonal lattice exhibits long time scale fluctuations. Top panels show the probability density of $\ell_{\nu} = L_{\nu}/4$ (see text)
        with the trajectories shown in the insets (the first 10\,ns is shown with a higher resolution to demonstrate anti-correlations). Although the symmetry axis flips between $x$, $y$ and $z$ at \SI{350}{\kelvin} (left) and \SI{365}{\kelvin} (middle), the system remembers its initial condition. At \SI{390}{\kelvin} (right) $\ell_x$, $\ell_y$ and $\ell_z$ are equivalent, although the distributions exhibit a fat tail. The lower panels show the joint probability densities for $2c/(a+b)$ (see text) and the potential energy per atom $u$. At \SI{350}{\kelvin} we see that the tetragonal lattice dominates, with the bcc phase becoming more significant at \SI{365}{\kelvin}. At \SI{390}{\kelvin}, the bcc structure dominates. The tetragonal structure is approx. 5\,meV/atom more stable than the bcc structure.
    }
\end{figure*}

%----------------------------------------------------------------------------------------------------------------
% ------------------- 13th paragraph -----------------------------------------------------------------------------
%----------------------------------------------------------------------------------------------------------------
Based on analysis of X-ray diffraction data, Burley proposed that differences 
in the \ce{Ag+} distributions in the high temperature phase play a determining role in the memory effect~\cite{Burley1967TheIodide}. 
Our simulation results potentially lend support to such a hypothesis.
To probe this notion further, for both wurtzite and zincblende as initial structures,
we generated an ensemble of eight independent trajectories of \SI{1}{\nano\second} at $T=350$\,K and $P=0$\,GPa,
and a further eight each at $T=390$\,K.  
Following this initial $NPT$ trajectory, $NVT$ simulations of \SI{1}{\nano\second} were performed using the average cell parameters, from which the structure factors, $S(|\vec{k}|)$, were calculated.
These are presented in \autoref{fig:sf}.

%----------------------------------------------------------------------------------------------------------------
% ------------------- 14th paragraph -----------------------------------------------------------------------------
%----------------------------------------------------------------------------------------------------------------
At \SI{350}{\kelvin} the peaks in $S(\vec{k})$ deviate
from the ideal tetragonal lattice owing to the persistent distortions in the \ce{I-} framework.
Of particular note is a ``super-structuring'' peak at 
$|\vec{k}| \approx 0.87$\,\AA$^{-1}$, 
which is indicative of long-range ordering of the \ce{Ag+} (see partial structure factors in Fig. S6~\cite{prl_sm}).
A trace of this super-structuring peak persists in the bcc phase at \SI{390}{\kelvin}, at least on the nanosecond time scales accessible in our simulations.
Importantly, similar to Burley's observations, the intensities of the peaks consistent with the bcc framework are different from the wurtzite- and zincblende-derived structures.
In addition, at $|\vec{k}| \approx 3.93$\,\AA$^{-1}$, corresponding to Miller indices [3, 1, 0],
there is a minor peak present for the wurtzite-derived structure that is absent in that derived 
from zincblende.
Thus, on the time and length scales accessible in our simulations, 
the bcc structures we observe retain memory of their low temperature parent phase.

%----------------------------------------------------------------------------------------------------------------
% ------------------- 16th paragraph -----------------------------------------------------------------------------
%----------------------------------------------------------------------------------------------------------------
% MD with random Ag config ------------------------------------------------------------------------
While Burley mentions splitting of the diffraction pattern near the transition 
temperature consistent with a tetragonal cell, we are not aware of clear experimental evidence of this structure. 
Therefore, to better understand the tet$^\prime$ structures and their relationship to the bcc phase, we performed a set of simulations with a smaller system size at $P=0$\,GPa and {$T/{\rm K}=350, 365, 390$}.
Specifically, we initialize the system with a $4\times 4 \times 4$ bcc iodide framework, and randomly distribute the \ce{Ag+}, while avoiding unphysical overlap.
Using this smaller system size allows us to sample over time scales exceeding \SI{100}{\nano\second}.
In \autoref{fig:rand}, we show histograms of $\ell_\nu = L_\nu / 4$, where $L_\nu$ is the simulation cell length along $\nu \in (x, y, z)$, along
with their respective time series shown in the insets.

%----------------------------------------------------------------------------------------------------------------
% ------------------- 17th paragraph -----------------------------------------------------------------------------
%----------------------------------------------------------------------------------------------------------------
At \SI{350}{\kelvin}, we immediately see that the system transforms into a tetragonal lattice,
with lattice parameters consistent with our earlier space group analysis.
However, we also see that the direction of the $c$-axis flips between $x$, $y$ and $z$, which corresponds to switching the mode of symmetry breaking in the \ce{Ag+} distribution (see Fig. S7~\cite{prl_sm}).
Given this observed flipping of the $c$-axis, we might expect the averaged structure to resemble, over sufficiently long times, the bcc lattice with $\ell_x = \ell_y = \ell_z$. 
However, the histograms presented in \autoref{fig:rand} indicate that the $c$-axis preferentially aligns along the $x$-axis.
Despite the computational efficiency of the MLIP compared to AIMD, probing longer time and length scales remains too computationally demanding, and we therefore cannot conclude definitively whether these observations correspond to: (i) a genuine breaking of ergodicity; or (ii) a relevant time scale that exceeds the length of our simulations.
While either could provide a mechanism for the memory effect, scenario (ii) seems in line with experimental diffraction patterns that show bcc symmetry, and the observation that the memory effect depends on kinetic factors such as the cooling rate~\cite{Burley1967TheIodide,Smith2019StackingPhase}.

%----------------------------------------------------------------------------------------------------------------
% ------------------- 18th paragraph -----------------------------------------------------------------------------
%----------------------------------------------------------------------------------------------------------------
For a closer analysis of these lattice structures,
we assign $(a, b, c) $ to sorted ${\{\ell_\nu\}}$ such that $a \le b \le c$ at every instance.
We can then quantify the degree of spontaneous tetragonal distortion by $2c/(a+b)$.
The joint probability density of $2c/(a+b)$ with the potential energy per atom $u$ is presented in \autoref{fig:rand}
which confirms that tetragonal lattices dominate at \SI{350}{\kelvin} and bcc lattices at \SI{390}{\kelvin}. 
At both temperatures, however, the probability densities indicate significant populations of the competing lattice structure.
For the intermediate temperature of \SI{365}{\kelvin}, both bcc and tetragonal lattices 
are observed with significant probabilities, with the tetragonal lattice lower in potential energy by approximately \SI{5}{\milli\electronvolt}/atom. 
This suggests that the bcc lattice is stabilized by entropic effects.
While confirming with appropriate finite size scaling remains too computationally demanding for this study, 
these observations are strongly suggestive of a first-order transition. 
(In the SM~\cite{prl_sm}, we present results from simulations where we bias the 
Ag--Ag interaction that supports this notion.)
This would further suggest a bcc-tet$^\prime$ boundary in the $\rm P$-$\rm T$ plane of the phase diagram.
Termination of such a boundary at a critical point offers a possible explanation for
the anomalous heat capacity, similar to the theory of Perrott and Fletcher~\cite{Perrott1968HeatTheory}.

%----------------------------------------------------------------------------------------------------------------
% ------------------- 19th paragraph -----------------------------------------------------------------------------
%----------------------------------------------------------------------------------------------------------------
In summary, our results suggest that for AgI at temperatures just above the superionic transition, 
a distorted tetragonal lattice of \ce{I-} competes with the bcc framework, which is accompanied by broken translational symmetry of the \ce{Ag+} ions. 
This finding offers a potential mechanism for the experimentally observed memory effect, and may be relevant for the anomalous heat capacity at high temperatures.
More broadly, order-disorder transitions are common in many ionic conductors~\cite{Gao2020ClassicalConductors} 
and whether these also underpin similar memory effects in other type I superionic conductors 
with stacking disordered low temperature phases remains an open question. 
While the machine learned representation of the PES that we employ is highly accurate, there are, of course, limitations on the accuracy of the underlying DFT functional (PBE+D3). 
For example, we observe a shift of the transition temperature in our simulations by approx. \SI{70}{\kelvin} compared to experiment, and it is possible that the pressure is similarly offset.
The experimental relevance of the intermediate structures observed just below the superionic transition temperature ($T\lesssim 330$\,K) is also unclear at ambient pressure, and remains the topic of future study.

%----------------------------------------------------------------------------------------------------------------
% ------------------- 20th paragraph -----------------------------------------------------------------------------
%----------------------------------------------------------------------------------------------------------------
Notwithstanding such shortcomings, our observations appear in line with many curious experimental observations.
The fine energy difference of a few meV/atom between the bcc and tetragonal structures suggests that, close to the transition temperature, both of these competing structures may be present in abundance.
Further studies, both computational and experimental, are needed, however, to establish whether this is in the form of phase coexistence, or long-lived thermal fluctuations. 
Any such future work, may also help to establish the relative abundance of these different structures, and shed light on the ionic transport mechanisms.

%----------------------------------------------------------------------------------------------------------------
% ------------------- ------------- -----------------------------------------------------------------------------
%----------------------------------------------------------------------------------------------------------------
\begin{acknowledgments}
We thank Christoph Salzmann for helpful discussions.
Access to CSD3 was obtained through a University of Cambridge EPSRC Core Equipment Award EP/X034712/1.
SJC is a Royal Society University Research Fellow at the University of Cambridge (URF\textbackslash R1\textbackslash 211144).
WB and CG thank the AFRL for partial funding of this project through grant FA8655-21-1-7010.
\end{acknowledgments}

\textbf{Conflicts of interest:} GC has equity stake in Symmetric Group LLP and \AA{}ngstrom AI Inc., companies engaged in the application of machine learning to material and molecular simulation.

%\FloatBarrier

%----------------------------------------------------------------------------------------------------------------

%Control: key (1)

\end{document}

\title{Supplemental Material for:\\Symmetry Breaking in the Superionic Phase of Silver-Iodide}

\date{\today}

\maketitle

\tableofcontents

%############################################################################################
%####################### Text  ##############################################################
%############################################################################################

\section{MACE Potential: Training and Tests}

The initial models are created from {100} configurations sampled from MD simulations using the empirical PRV force-field.
An active learning scheme with a committee \cite{Schran2020} of five MACE models, trained with different random seeds, is used for sampling the training data.
A standard deviation of 100\,meV/\AA{} in the predicted forces of the committee is used as the threshold for sampling new training configurations.
A training task with a specific initial configuration, at a given temperature and pressure, is considered concluded if the committee disagreement remains below the threshold for at least \SI{200}{\pico\second} of continuous isobaric-isothermal MD simulation. Discussion of this training method can be found in Ref~\cite{Schran2020}.

The main focus of this work is the $\beta/\gamma \to \alpha$ transition at a pressure of 0\,\si{\giga\pascal}.
However, we wanted the model to also be transferable to other significant phases which are experimentally observed at higher pressures (e.g. rocksalt). 
We therefore ensured that the settings we used for the DFT calculations gave converged results across all known phases.
Therefore, in addition to initializing trajectories with wurtzite, zincblende, and rocksalt structures, all other AgI crystal structures available in the \emph{Materials Project} online repository~\cite{Jain2013Commentary:Innovation} are also utilized for training.
Transferability to the rocksalt phase required us to utilize a higher k-point density and a small Gaussian smearing to converge the self-consistent electronic structure calculations.
Convergence of the self-consistent field (SCF) calculations
across different phases is achieved by application of
a plane wave cutoff of \SI{70}{\rydberg},
k-point density of \SI{3.25}{\angstrom^{-1}},
Gaussian smearing of \SI{0.005}{\rydberg},
and a SCF convergence threshold of $10^{-8}$\,\si{\rydberg}.

Training is carried out most extensively at pressures of 0\,GPa (0.1\,GPa for rocksalt) and 1\,\si{\giga\pascal} and at every \SI{50}{\kelvin} within the temperature range $250 \lesssim T/{\rm K} \lesssim 500$.
At 0\,\si{\giga\pascal}, higher temperatures of up to \SI{800}{\kelvin} in \SI{100}{\kelvin} intervals are also surveyed, but full exploration of the molten phases is not guaranteed. 
In total, 1704/740 training/test configurations are sampled.
Despite including these higher pressure structures, complete transferability across the full phase diagram will likely require further training.
A model, chosen randomly from the committee to represent the PES, demonstrated an overall accuracy with a root mean squared error (RMSE) of ($2$\,meV/atom, $15$\,meV/\AA{}) for (energy, force) predictions.
These RMSE values, however, include contributions from trajectories with $T>500$\,K. 
In the temperature range we are interested in for this work, $270\leq T/{\rm K}\leq 450$, 
the model exhibits an improved accuracy, with an RMSE of less than ($1$\,meV/atom, $13$\,meV/\AA).
The training data and the models are made available online~\footnote{\url{https://gitlab.com/amirhajibabaei/bulkagi.git}}.

\begin{figure*}
    \centering
    \includegraphics[width=0.95\textwidth]{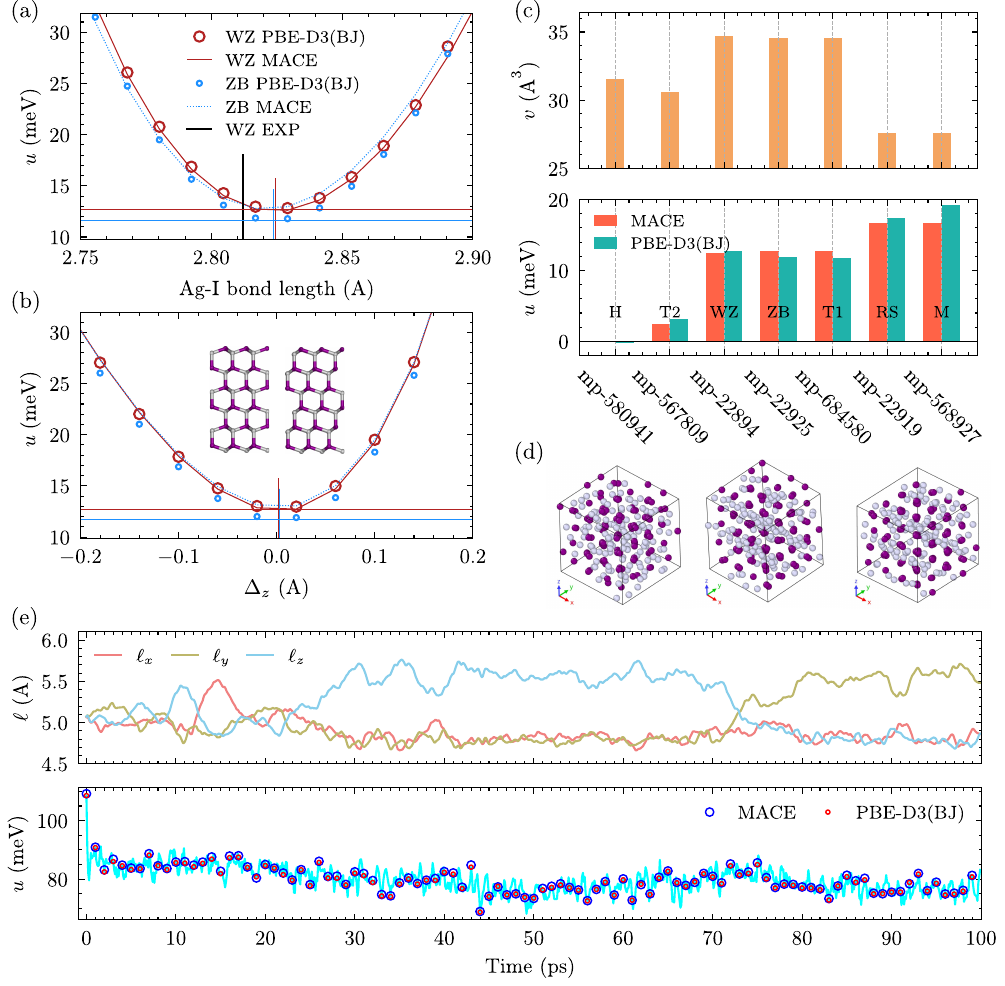}
    \caption{
    \label{fig:mace-validation}
    Validation of the MACE potential.
    Potential energy per atom of the wurtzite and zincblende lattices are shown in (a) as a function of Ag-I bond length
    and in (b) as a function of \ce{I-} sublattice displacement along the $c$-axis. 
    (c) The potential energies, after optimization with the trained MACE model, of crystal structures taken from the \emph{Materials Project} online database.
    {(d) Snapshots taken at 0, 50, and \SI{100}{\pico\second} from isobaric-isothermal MD at \SI{350}{\kelvin} and \SI{0}{\giga\pascal}, starting from a bcc \ce{I-} lattice and a random \ce{Ag+} configuration, and (e) the potential energy per atom $u$ and $\ell_\nu$ vs time. Energies calculated with PBE+D3 for configurations resampled from this trajectory are shown by red symbols, and confirm the accuracy of the MACE model.} 
    }    
\end{figure*}

We further evaluated the model's accuracy against the experimental data available for the lattice constants.
A wurtzite lattice is often described using three parameters $a$, $c$, and $u$, which are related to the Ag-I bond lengths via:
\begin{equation}
    \begin{split}
        b_1 &= c \sqrt{\frac{1}{3}\left(\frac{a}{c}\right)^2 + \left(\frac{1}{2} - u\right)^2} \\
        b_2 &= cu
    \end{split}
\end{equation}
where three $b_1$ bonds and a single $b_2$ bond form a tetrahedron.
The experimental lattice parameters $a = 4.59$, $c = 7.51$, and $u = 0.377$, calculated in Ref~\cite{Burley1963StructureIodide}, are consistent with nearly perfect tetrahedral structures where $c/a = \sqrt{8/3}$, $u = 3/8$, and $b_1 = b_2 = b$.
The same applies to the zincblende structure, as the only difference is the stacking order along the polar $c$-axis.
Thus, the optimal lattice for both cases can be described using only the Ag-I bond length, as shown in \autoref{fig:mace-validation}(a).
To investigate the distortion of the tetrahedra, we calculated the potential energy as a function of distortion along the $c$-axis while keeping the other lattice parameters $a$ and $c$ fixed.
In this context, distortion is described by $\Delta_z$, which is the displacement of the I sublattice along the $c$-axis relative to the perfect tetrahedral position.
The potential energy profile shown in \autoref{fig:mace-validation}(b) confirms that the ideal tetrahedral coordination is indeed optimal.

Aside from wurtzite and zincblende lattices, other AgI crystal structures
listed in the \emph{materials-project} online repository~\cite{Jain2013Commentary:Innovation},
are also used for testing the MACE potential by calculating
the optimized lattice parameters.
First, a full (positions+cell) optimization is carried out
using the MACE potential with a convergence criterion of
$|f_{max}| < \SI{1}{\milli\electronvolt\per\angstrom}$.
Then, single-point DFT energies and forces are calculated and
compared to those obtained with MACE.
As shown in \autoref{fig:mace-validation} (c), the energies are in excellent agreement. 
%\tcb{
The superionic bcc phase is excluded from this relaxation analysis because it is considered an entropically stabilized lattice~\cite{Mendels2018SearchingIodide}.
%}

An important aspect of this work is the emergence of a distorted tetragonal \ce{I-} framework over a range of temperatures instead of the expected bcc lattice.
Although we were initially unaware of this structure, {it was included in the training as a derivative structure} during the active learning algorithm. 
To further confirm the accuracy of the MACE model, we conducted an isothermal-isobaric MD simulation at \SI{350}{\kelvin}.
In this simulation, a bcc \ce{I-} framework with a random \ce{Ag+} distribution transformed into a tetragonal lattice.
We tested the predicted energy and forces against the underlying PBE-D3 at \SI{1}{\pico\second} intervals to validate the accuracy of the MACE model.
The accuracy of the model is demonstrated in \autoref{fig:mace-validation}(e) and is quantified by an RMSE of ($0.4$\,meV/atom, $13$\,meV/\AA) for (energy, force) predictions.

In addition, for the trajectories used to generate results presented in Fig.~1 of the main text, we resampled 100 snapshots (separated by 10\,ps) from the final 1\,ns, and recomputed the energy and forces among all committee members used during the active learning step.
For a measure of spread in the committee predictions, we define:
\begin{gather} \label{eq:spread}
    \Delta (y) = \max_k |y_k - \overline{y}| \\
    \overline{y} = \frac{1}{n_c}\sum_{k=1}^{n_c} y_k
\end{gather}
where $y$ is an energy or force value, $k$ refers to the committee members, and $n_c=5$ in the size of committee.
The maximum spreads in the predicted energy and forces of the resampled configurations are presented in Table~\ref{tab:committee}.

\begin{table}[]
     \centering
     \caption{
        The maximum spreads (see \autoref{eq:spread}) in the committee predictions in 100 configurations resampled from each trajectory.
     }
     \label{tab:committee}
     \bgroup
         \setlength\tabcolsep{10pt}
         \begin{tabular}{c|cc|cc}
         \toprule
           &  \multicolumn{2}{c}{wz$^\ast$}  &  \multicolumn{2}{c}{zb$^\ast$} \\
             Temperature & Energy & Force &  Energy & Force \\
              (\si{\kelvin}) & (meV/atom) & (meV/\AA{}) & (meV/atom) & (meV/\AA{}) \\
            \midrule
              270 & 0.08 & 20.89 & 0.25 & 21.91 \\
              320 & 0.35 & 29.32 & 0.39 & 39.10 \\
              350 & 0.13 & 37.12 & 0.15 & 38.67 \\
              390 & 0.14 & 44.68 & 0.14 & 48.91 \\
              450 & 0.13 & 70.06 & 0.14 & 47.62 \\
         \bottomrule
         \end{tabular}
     \egroup
\end{table}

\begin{figure*}
    \centering
    \includegraphics[width=0.95\textwidth]{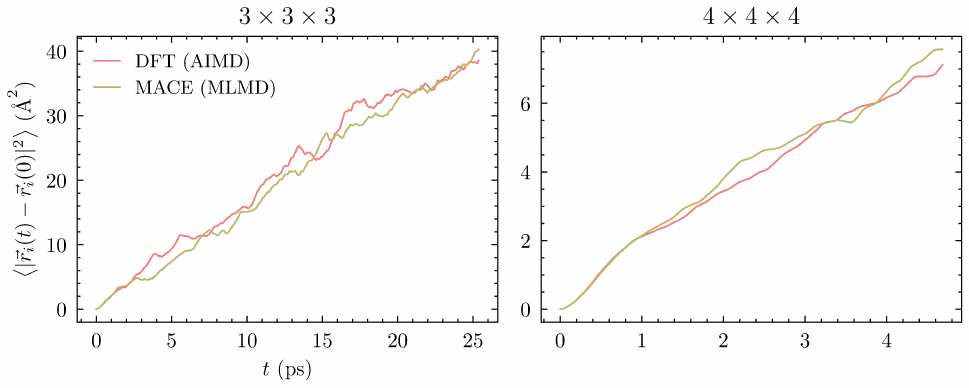}
    \caption{
    \label{fig:aimd}
    Short-time \ce{Ag+} MSDs from identical initial configurations using both DFT and the MACE MLIP, with MD simulations in the $NVE$ ensemble.
    Initial configurations are obtained from equilibration of systems with a random \ce{Ag+} distribution and either a $3\times 3 \times 3$ (left) 
    and $4\times 4\times 4$ (right) bcc \ce{I-} framework, at a temperature of \SI{450}{K} in the $NVT$ ensemble. $\langle\cdot\rangle$ represents averaging over \ce{Ag+} ions.
    }
\end{figure*}

To validate the accuracy of the MACE potential for dynamics, 
we compared \ce{Ag+} MSDs with those obtained from short AIMD simulations.
Initial configurations were generated by equilibrating structures with a bcc \ce{I-} framework and a random \ce{Ag+} distribution in the $NVT$ ensemble
at $T=\SI{450}{\kelvin}$.
These configurations were then used for short MD runs in the $NVE$ ensemble using the Verlet integration algorithm with both MACE and DFT. 
To enhance the performance of DFT calculations, we reduced the SCF convergence threshold to $10^{-6}$\,\si{\rydberg} and employed a $\Gamma$-only k-point grid. 
The resulting MSDs, shown in \autoref{fig:aimd}, confirm the accuracy of the MACE potential.

\section{Space Group Analysis of the Iodine Framework \label{sec:spg}}

\begin{figure*}
    \centering
    \includegraphics[scale=0.8]{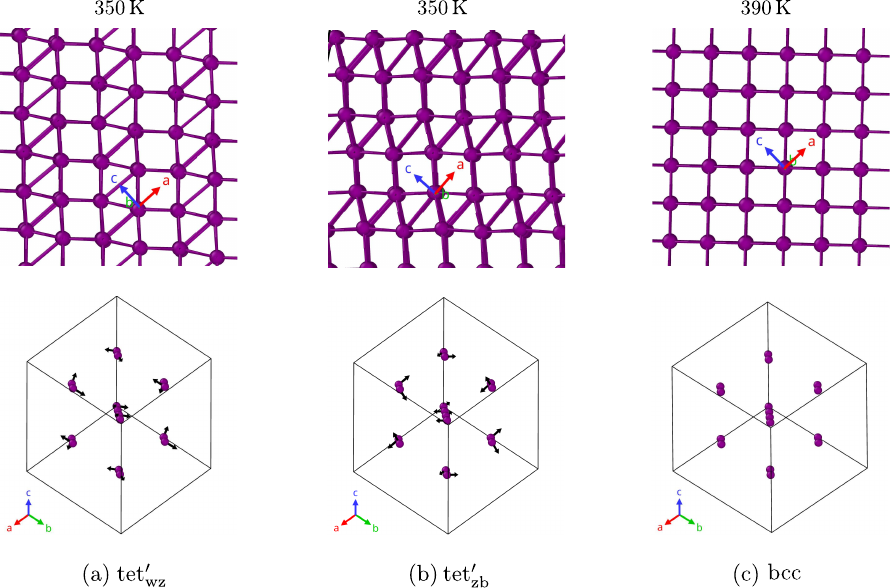}
    \caption{
    \label{fig:iodine-framework}
    The average \ce{I-} framework at temperatures corresponding to (a, b) tet$'$ and (c) bcc phases. 
    In the lower panels of (a, b),
    average displacements vectors from ideal tetragonal sites are shown in the {(parent phase dependent) tet$^\prime$ phases}.
    All of these displacement vectors have nearly the same length (\SI{0.26\pm0.01}{\angstrom}); although scaled by 4x for better visibility.}
\end{figure*}

Except for the intermediate phase tet$^\prime$, where the lattice is distorted, space group analysis of the average \ce{I-} framework is straightforward using the \texttt{spglib} software~\cite{Togo2018textttSpglib:Search}.
The distortions in the intermediate phase, as demonstrated in \autoref{fig:iodine-framework}, make its space group analysis nontrivial.
Furthermore, it seems that the distortions (in e.g. tet$^\prime_{\rm wz}$/tet$^\prime_{\rm zb}$) depend on the parent configurations (e.g. wurtzite/zincblende) which are used for initializing the MD simulations.
It is worth noting that usually a transient bcc lattice is observed before the distorted tetragonal phase.
Because of the relatively large sizes (1539 atoms), after initial deformations, the cell fluctuations remained negligible
for the approximately \SI{3}{\nano\second} long MD simulations in the $NPT$ ensemble.
Subsequently, a \SI{6}{\nano\second}/\SI{4}{\nano\second} simulations in the $NVT$ ensemble at temperatures \SI{350}{\kelvin}/\SI{390}{\kelvin} are performed to calculate the average positions of \ce{I-} anions as well as \ce{Ag+} distribution which is further discussed in \autoref{sec:cations}.

The following algorithm is used for analyzing the distorted \ce{I-} framework at the temperature of \SI{350}{\kelvin}:
\begin{enumerate}
    \item \textbf{Clustering into Sublattices}:
    The \ce{I-} framework is divided into sublattices using a naive clustering algorithm that employs nearest neighbor vectors, defined by Voronoi constructions, as a measure of similarity. This approach detected 8 sublattices (each with 2 atoms per standard cell) with similar nearest neighbor vectors.

    \item \textbf{Space Group Analysis of Sublattices}:
    The space group of each sublattice is analyzed using \texttt{spglib}. 
    All sublattices were found to have the same space group I4/mmm (number 139) and nearly identical lattice parameters.
    These parameters, combined with the relative shifts of the sublattices, are used to construct a repeat unit for the distorted \ce{I-} framework.

    \item \textbf{Construction of the Repeat Unit}:
    The space group analysis of the sublattices indicated that the constructed repeat unit should be closely related to a body-centered tetragonal lattice.
    Considering the repeat unit as a $(2 \times 2 \times 2)$ super-structure, all of the atoms in the repeat unit were displaced from the ideal tetragonal sites by nearly the same length of \SI{0.26\pm0.01}{\angstrom}; but in varying directions as demonstrated in \autoref{fig:iodine-framework} (a) and (b).

    \item \textbf{Verification by Reconstruction}:
    The above analysis is verified by reconstructing the \ce{I-} framework from the repeat unit and the transformation matrix obtained from the earlier space group analysis.
    The \texttt{make-supercell} function available in the ASE Python package~\cite{HjorthLarsen2017TheAtoms} is used for reconstructing the lattice.
    The (average, maximum) difference between our reconstructed lattice and that directly obtained from MD was (0.07, 0.17)\,\si{\angstrom} for the wurtzite-derived lattice and (0.02, 0.043)\,\si{\angstrom} for the zincblende-derived lattice.
\end{enumerate}

It is also worth noting that a smaller repeat unit based on the C2/c or C2/m space groups {could in principle} provide a more precise description of the distorted lattices observed in the aforementioned instances.
However, the details of the distortions are subject to interference due to the long time scales of \ce{Ag+} density fluctuations between modes of symmetry breaking.
Consequently, the long-range distortions of the average \ce{I-} framework in the tet$^\prime$ phase can be sufficiently described using the $(2 \times 2 \times 2)$ tetragonal repeat unit.

\section{Cation Distribution\label{sec:cations}}

Having detected the repeat unit of the \ce{I-} framework in the superionic conducting phases, we investigated the \ce{Ag+} distribution observed in the $NVT$ trajectories described in \autoref{sec:spg}. 
As discussed in the previous section, the \ce{I-} framework at \SI{350}{\kelvin} can be described using a $(2\times2\times2)$ tetragonal repeat unit. We use the same repeat unit to describe the \ce{Ag+} distribution. A translation aside, the repeat unit vectors $\vec{t}_\nu$ for $\nu\in(1, 2, 3)$ are defined by:

\begin{equation}
    (\vec{t}_1 , \vec{t}_2 , \vec{t}_3)
    = 
    \mathbf{R}
    \begin{pmatrix}
        2a & 0  & 0  \\
        0  & 2a & 0  \\
        0  & 0  & 2c \\
    \end{pmatrix}
\end{equation}

where $\mathbf{R}$ is the rotation matrix obtained from the spacegroup analysis of the \ce{I-} framework.
Consequently, \ce{Ag+} positions are expressed as

\begin{equation}
    \vec{r} = \sum_{\nu=1}^{3} n_{\nu} \vec{t}_\nu + \vec{s}
\end{equation}
where the integers $n_{\nu}$ are chosen such that $\vec{s}$ is wrapped within the repeat unit.
A histogram of the vectors $\vec{s}$ is collected on a 3d grid with spacings of approximately \SI{0.2}{\angstrom} and is used to represent
the \ce{Ag+} distribution.
At \SI{390}{\kelvin}, although the average \ce{I-} lattice is bcc, 
we also use a $(2\times2\times2)$ repeat unit
for \ce{Ag+} distribution to fairly compare differences with the distribution observed at \SI{350}{\kelvin}. Slices of the obtained histograms are shown in the ${ab}$ plane along the $c$-axis in Fig. 1 of the main text.

\begin{figure*}
    \centering
    \includegraphics[scale=0.7]{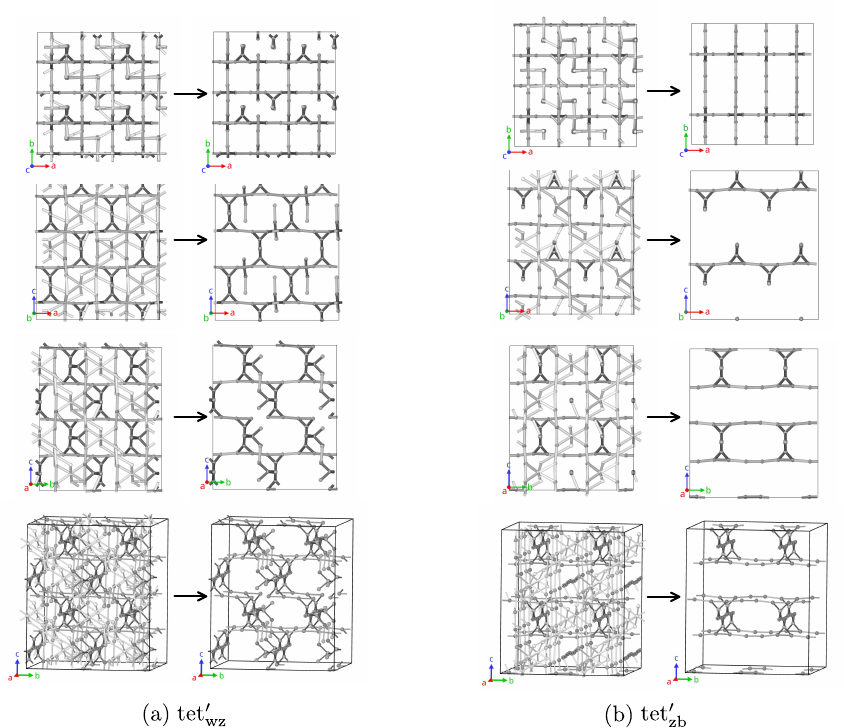}
    \caption{
    \label{fig:silver-distribution}
    The local peaks of the \ce{Ag+} density (shown in Fig.~1c of the main text) are shown for (a) tet$^\prime_{\rm wz}$ and (b) tet$^\prime_{\rm zb}$ structures from different angles.
    A shading is utilized to indicate prominence of the peaks (darker for taller) and the edges (darker for shorter length).
    While on the left hand side of the arrows all of the peaks are shown, on the right hand side only the largest cluster of the peaks with a distance cutoff of \SI{2.6}{\angstrom} are shown to clearly demonstrate the differences between tet$^\prime_{\rm wz}$ and tet$^\prime_{\rm zb}$.
    This demonstrates that the translational symmetry of the \ce{Ag+} distribution is not only broken, but also depends on the parent phase.
    }
\end{figure*}

For a site occupation analysis, we searched for the peaks of the \ce{Ag+} density.
The approximate locations of the peaks, with a resolution limited by the bin width, could be easily found by comparing the values on the grid.
For further refinement, the local density within \SI{1.0}{\angstrom} around each approximate maximum was interpolated by a cubic spline, and the refined location of the maximum was found using a local optimization algorithm.
The network of the peaks are shown in \autoref {fig:silver-distribution}.
We also show the largest cluster formed with a distance threshold of $d < \SI{2.6}{\angstrom}$ for the edges.
As demonstrated in \autoref{fig:silver-distribution}, this largest cluster contains the most prominent peaks as well.

\begin{figure*}
    \centering
    \includegraphics[scale=0.95]{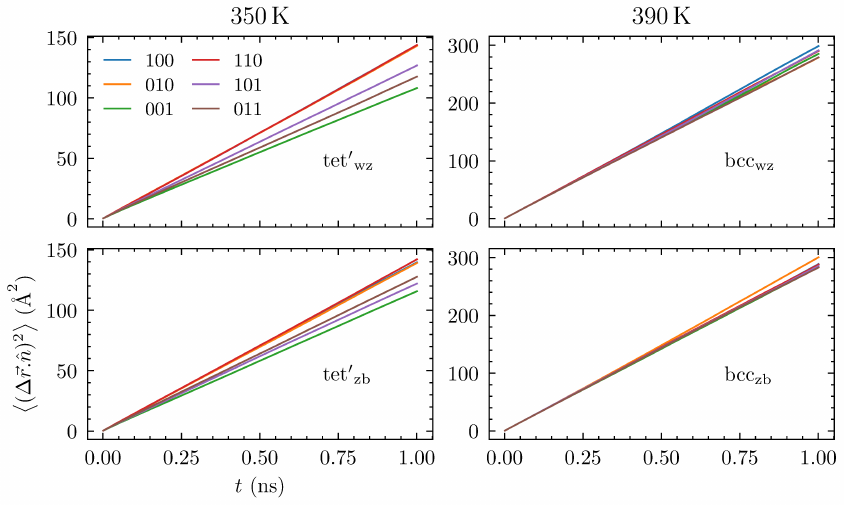}
    \caption{
    \label{fig:msd_aniso}
    Mean squared displacements along key crystallographic directions in the tet$^\prime$ (\SI{350}{\kelvin}) and bcc (\SI{390}{\kelvin}) phases, derived from wurtzite and zincblende parent structures.
    }
\end{figure*}

In \autoref{fig:msd_aniso}, we present preliminary results concerning diffusion in the distorted tetragonal phases. Specifically, we present
mean squared displacements along key crystallographic directions, demonstrating that diffusivity in the tet$^\prime$ phase is anisotropic and lower along the $c$-axis. 
A subtle difference also exists between the tet$^\prime_{\text{wz}}$ and tet$^\prime_{\text{zb}}$ phases.
In the former, diffusivity is slightly higher along the 101 direction than along 011, whereas in the latter the opposite is observed.

\section{Partial Structure Factors}

The structure factors (SFs) presented here are calculated using the \texttt{freud} package~\cite{Ramasubramani2020Freud:Data}.
Within this package, the partial SFs for two groups of atoms $\alpha,\beta$ are defined as:
\begin{equation}
S_{\alpha\beta}(\vec{k}) = \frac{1}{N} \sum_{i \in \alpha} \sum_{j \in \beta} e^{i\vec{k} \cdot \vec{r}_{ij}},
\end{equation}
which relate to the total SF by:
\begin{equation}
S(\vec{k}) - 1 = \sum_\alpha \sum_\beta \frac{N_\alpha N_\beta}{N_{\text{tot}}^2}(S_{\alpha\beta}(\vec{k}) - 1).
\end{equation}

\begin{figure*}
    \centering
    \includegraphics[width=0.95\textwidth]{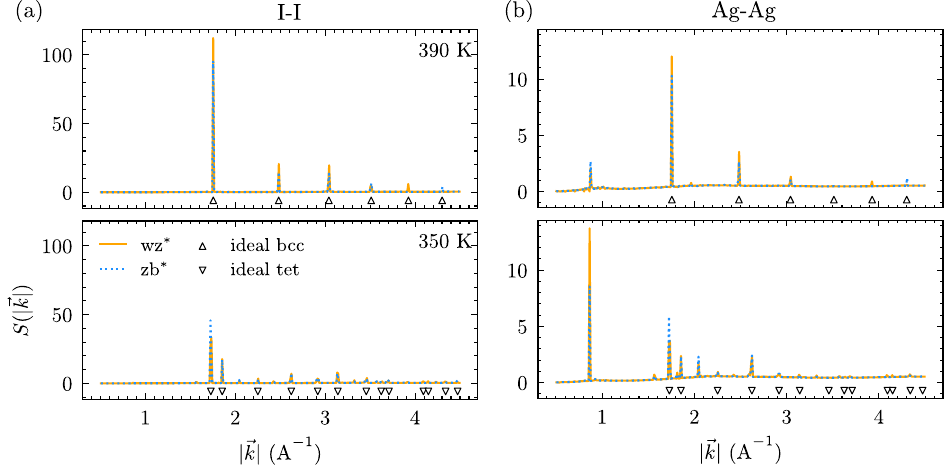}
    \caption{
    \label{fig:structure-factor}
    Partial structure factors at $T/K=350,390$ corresponding to tet$^\prime$ and bcc phases.
    The corresponding total structure factors are shown in Fig.~2 of the main article.
    This shows that the ``super-structuring'' peak at $|\vec{k}|=\SI{0.87}{\per\angstrom}$ is derived from Ag-Ag correlations.
    }
\end{figure*}

The partial SFs, contributing to the total SFs presented in Fig. 2 of the main text, are shown in \autoref{fig:structure-factor}.
These demonstrate that the super-structuring peak at the wave vector $|\Vec{k}|\approx 0.87 \si{\per\angstrom}$ is caused by ordering of \ce{Ag+} ions.
Importantly, this is the most dominant peak of the \ce{Ag+} partial SF at \SI{350}{\kelvin} corresponding to the tet$^\prime$ phase.
Persistence of this peak at \SI{390}{\kelvin} signals substantial presence of tet$^\prime$-like fluctuations within the bcc phase
since the corresponding wave vector is not commensurate with the bcc lattice.
Furthermore, with the tetragonal lattice parameters given by $a \approx 4.79$\,\AA{} and $c\approx 5.58$\,\AA{},
it is clear that the wave vectors contributing to the super-structuring peak are among the following
\begin{equation}\label{eq:sspeaks}
    \vec{k} = 
    \begin{cases}
        (\pm \frac{\pi}{a}, 0, \pm \frac{\pi}{c})    \\
        (0, \pm \frac{\pi}{a}, \pm \frac{\pi}{c})    
    \end{cases}
    \text{,}
\end{equation}
which correspond to the fractional Miller indices $q={[1/2, 0, 1/2]} $ and ${[0,1/2,1/2]}$.

\begin{figure*}
    \centering
    \includegraphics[width=0.95\textwidth]{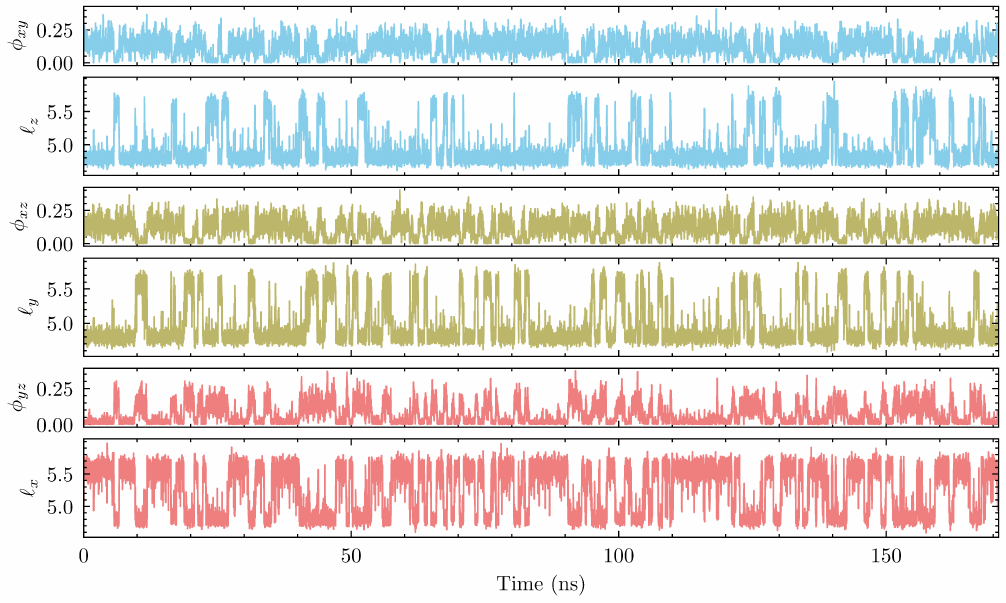}
    \caption{
    \label{fig:order-param}
    Switching of the mode of symmetry breaking in \ce{Ag+} distribution by flipping of the tetragonal $c$-axis.
    Trajectories are from MD simulations at \SI{350}{\kelvin} shown in Fig.~3 of the main article.
    }
\end{figure*}

For the MD simulations with $4\times 4\times 4$ bcc/tetragonal supercells presented in Fig. 3 of the main text, 
because of strong cell fluctuations in the $NPT$ ensemble,
we utilize the following quantities in lieu of the SF for probing the symmetry breaking in \ce{Ag+} distribution:
\begin{equation}
    \phi_{xy} = \bigg|\frac{1}{N} \sum_{i \in Ag} e^{i \pi \left(\frac{x_i}{\ell_x}+\frac{y_i}{\ell_y}\right)}\bigg|^2
               +\bigg|\frac{1}{N} \sum_{i \in Ag} e^{i \pi \left(\frac{x_i}{\ell_x}-\frac{y_i}{\ell_y}\right)}\bigg|^2,
\end{equation}
where $\ell_\nu = \frac{1}{4}L_\nu$ for $\nu$ in $(x, y, z)$, with $\phi_{xz}$ and $\phi_{yz}$ defined similarly.
From the trajectories presented in \autoref{fig:order-param} for the simulations at \SI{350}{\kelvin},
the frequent flipping of the tetragonal $c$-axis between the $x$, $y$ and $z$ directions
is always mirrored by the simultaneous changes in $\phi_{xy}$, $\phi_{yz}$, and $\phi_{xz}$
indicating a corresponding mode switching in the ordering of \ce{Ag+} cations.
Consistent with \autoref{eq:sspeaks}, the tetragonal $c$-axis always remains orthogonal to the plane $\mu\nu$ for which $\phi_{\mu\nu} \approx 0$.
Thus, the flipping of the $c$-axis is strongly correlated with the switching of symmetry-breaking modes in the \ce{Ag+} site occupations.

\section{Survival Analysis of the Superionic Phases}

\begin{figure*}
    \centering
    \includegraphics[width=0.95\textwidth]{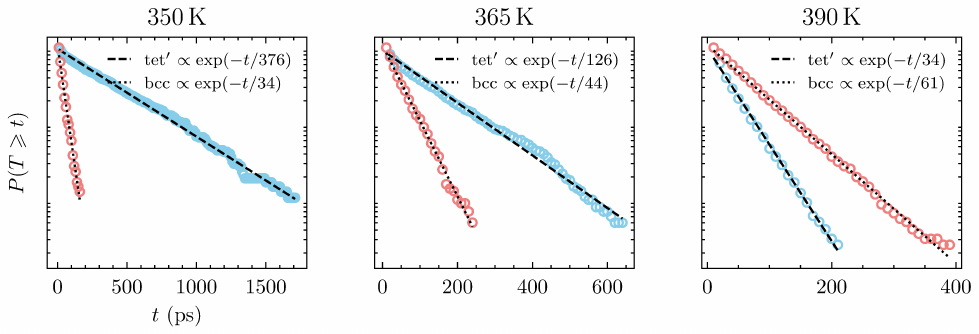}
    \caption{
    \label{fig:survival}
    Survival time scales of the tet$^\prime$ and bcc phases in $NPT$ simulations of systems with a $4 \times 4 \times 4$ bcc \ce{I^-} framework.
    $P(T \geqslant t)$ is the survival function representing the probability that a single phase will last for a time $T$ longer than $t$.
    }
\end{figure*}

In Fig.~3 of the main text, dynamics of the system with a $4\times 4 \times 4$ bcc \ce{I-} framework
is shown in the $NPT$ ensemble demonstrating that the system successively transitions between bcc and tet$^\prime$ states.
At lower temperatures, the tet$^\prime$ phase dominates, while at higher temperatures, the bcc phase becomes more prominent. 
However, strong fluctuations towards the competing phase persist. 
The tet$^\prime$ phase can occur with $c$-axis along $(x, y, z)$ directions and flipping of the direction of the $c$-axis 
occurs through a transient bcc phase.
The long-time trajectories obtained for this system enable the extraction of a statistical distribution of time scales, with $p(t)$ representing the probability that a phase lasts for a time $t$,
from which we calculate the survival function $P(T \geqslant t)=\int_t^\infty p(t^\prime) dt^\prime$, defined as the probability that the lifetime $T$ is larger than $t$.
As shown in \autoref{fig:survival}, the survival function follows a functional form $\exp\left( -t/\tau \right)$, where $\tau$ is the characteristic time scale.
This leads to the corresponding probability distribution function $p(t) = \frac{1}{\tau} \exp\left( -t/\tau \right)$. 
The characteristic timescales obtained from this analysis are reported directly in the figure legend. 
Note that these timescales will likely depend on the system size, though probing this in detail lies beyond the scope of the present study.

\section{Phase Switching by Biasing Cation Interactions}

The persistence of the super-structuring peak in the structure factor at \SI{390}{\kelvin} demonstrates the presence of strong tet$'$-like fluctuations within the bcc phase.
The MD simulations presented in Fig. 3 of the main text suggest that these fluctuations might manifest in a manner akin to tet$'$-bcc coexistence.
However, given the fine potential energy difference of approximately 5 meV/atom, further confirmation through finite size scaling is required to rule out long-lived distortions instead of phase coexistence.

\begin{figure*}
    \centering
    \includegraphics[width=0.95\textwidth]{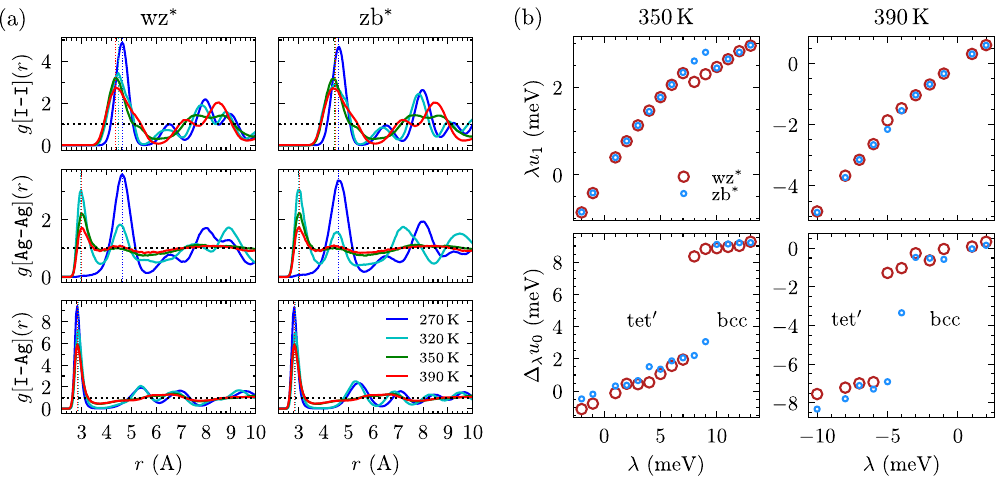}
    \caption{
    \label{fig:radial-distributions}
    (a) Radial distribution functions (RDFs) of the wurtzite- and zincblende-derived phases at representative temperatures.
    (b) The changes of the potential energy per atom as a function of applied repulsive ($\lambda >0 $) and attractive ($\lambda < 0 $) Ag-Ag biasing pair potentials, respectively, at \SI{350}{\kelvin} and \SI{390}{\kelvin}.
    }
\end{figure*}

As an alternative to finite size scaling which is prohibitive because of the long characteristic time scales,
we consider a perturbation of the PES, defined as follows
\begin{gather}
    U_\lambda = U_0 + \lambda U_1 \\
    U_1 = \frac{1}{2} \sum_{i>j} \left(1 - \tanh \left(\frac{r_{ij} - r_0}{\sigma}\right)\right)
\end{gather}
where the summation over $i, j$ includes all \ce{Ag+} cations.
The constants $r_0 = 3$\,\si{\angstrom} and $\sigma = 0.2$\,\si{\angstrom} are chosen 
based on the first peak of the Ag-Ag RDF (see \autoref{fig:radial-distributions}).
We utilize the earlier wurtzite- and zincblende-derived superionic configurations at temperatures of \SI{350}{\kelvin} and \SI{390}{\kelvin}, corresponding to the tet$^\prime$ and bcc phases, respectively.
Molecular dynamics simulations are then performed for \SI{1}{\nano\second} in the $NPT$ ensemble with varying degrees of perturbation $\lambda$.

The variations in the potential energy per atom $\Delta_{\lambda} u_0$ 
{(i.e., not including the contribution from the bias itself)} as a function of the perturbation scale $\lambda$ are depicted in \autoref{fig:radial-distributions}.
It is observed that a small degree of repulsive interactions between \ce{Ag+} ions induces a transformation from the tet$^\prime$ phase at \SI{350}{\kelvin} to the bcc phase. 
Conversely, a small degree of attractive interactions results in the transformation from the bcc phase at \SI{390}{\kelvin} to the tet$^\prime$ phase.
The internal energies of the tet$^\prime$ and bcc phases, estimated from the gap in $\Delta_{\lambda} u_0$, 
is in good agreement with the results shown in Fig.~3 of the main article.

\section{Diffusivity in Non-Superionic Phases}

\begin{figure}
    \centering
    \includegraphics[]{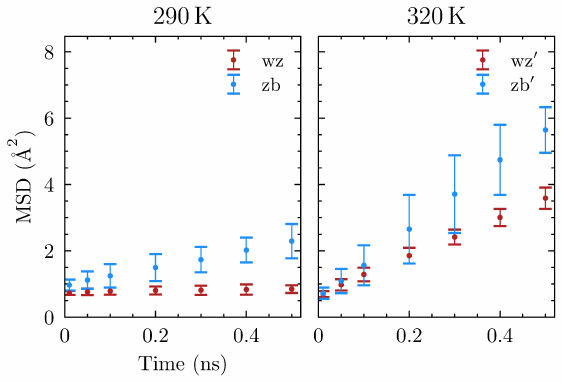}
    \caption{
    \label{fig:nonsuperionic_msds}
    \ce{Ag+} MSDs at selective temperatures below the superionic transition.
    }
\end{figure}

\autoref{fig:nonsuperionic_msds} presents the mean square displacement (MSD) of \ce{Ag+} cations obtained from simulations at two different temperatures. 
At \SI{290}{\kelvin}, where both the wurtzite (wz) and zincblende (zb) structures remain stable, no significant diffusion is observed in the wurtzite phase, while only slight diffusion is evident in the zincblende phase. 
At \SI{320}{\kelvin}, following the redistribution of \ce{Ag+} ions, a significant increase in \ce{Ag+} diffusivity is observed in both phases.
These phases are distinguished from the ideal wurtzite and zincblende structures by a prime superscript i.e. by wz$^\prime$ and zb$^\prime$ respectively.

\FloatBarrier
%############################################################################################
%####################### Figures ############################################################
%############################################################################################

